\pdfoutput=1
\documentclass[11pt]{article}

\usepackage{ACL2023}

\usepackage{times}
\usepackage{latexsym}
\usepackage{hyperref}

\usepackage[T1]{fontenc}
\usepackage[utf8]{inputenc}

\usepackage{microtype}

\usepackage{inconsolata}

\usepackage{longtable}

\usepackage{xcolor}
\usepackage{colortbl}
\usepackage{multicol}
\usepackage{booktabs}
\usepackage{makecell}
\usepackage{amsmath, amssymb}
\usepackage{multirow}
\usepackage{mathtools}
\usepackage{enumitem}
\usepackage{subcaption}
\usepackage{float}
\usepackage{graphicx}
\usepackage{caption} 
\usepackage{upgreek}
\usepackage{seqsplit}
\usepackage{color,soul}
\usepackage{arydshln}

\usepackage{amsmath,amsfonts,bm}

\def\eqref#1{equation~\ref{#1}}
\def\1{\bm{1}}

\def\vx{{\bm{x}}}

\DeclareMathAlphabet{\mathsfit}{\encodingdefault}{\sfdefault}{m}{sl}
\SetMathAlphabet{\mathsfit}{bold}{\encodingdefault}{\sfdefault}{bx}{n}

\DeclareMathOperator*{\argmax}{arg\,max}

\usepackage{caption}
\usepackage{subcaption}
\definecolor{gg}{gray}{0.92}
\newcolumntype{a}{>{\columncolor{gg}}c}

\definecolor{lightgreen}{rgb}{0.56, 0.93, 0.56}

\DeclareRobustCommand{\hlyellow}[1]{{\sethlcolor{yellow}\hl{#1}}}
\DeclareRobustCommand{\hlgreen}[1]{{\sethlcolor{lightgreen}\hl{#1}}}

\definecolor{realblue}{RGB}{0,0,255}

\title{Direct Fact Retrieval from Knowledge Graphs without Entity Linking}

\author{
    Jinheon Baek$^1$\thanks{\hspace{0.2cm} Work done while interning at Amazon. Corresponding author: Jinheon Baek (\texttt{jinheon.baek@kaist.ac.kr})} \quad
    \quad Alham Fikri Aji$^2$ \quad
    \quad Jens Lehmann$^{3}$ \quad
    \quad Sung Ju Hwang$^1$ \\
    KAIST$^{1}$ \quad MBZUAI$^{2}$ \quad Amazon$^{3}$ \\
    \texttt{\{jinheon.baek, sjhwang82\}@kaist.ac.kr} \\
    \texttt{alham.fikri@mbzuai.ac.ae} \quad \texttt{jlehmnn@amazon.com}
}

\begin{document}
\maketitle

\begin{abstract}
There has been a surge of interest in utilizing Knowledge Graphs (KGs) for various natural language processing/understanding tasks. The conventional mechanism to retrieve facts in KGs usually involves three steps: entity span detection, entity disambiguation, and relation classification. However, this approach requires additional labels for training each of the three subcomponents in addition to pairs of input texts and facts, and also may accumulate errors propagated from failures in previous steps. To tackle these limitations, we propose a simple knowledge retrieval framework, which directly retrieves facts from the KGs given the input text based on their representational similarities, which we refer to as Direct Fact Retrieval (DiFaR). Specifically, we first embed all facts in KGs onto a dense embedding space by using a language model trained by only pairs of input texts and facts, and then provide the nearest facts in response to the input text. Since the fact, consisting of only two entities and one relation, has little context to encode, we propose to further refine ranks of top-$k$ retrieved facts with a reranker that contextualizes the input text and the fact jointly. We validate our DiFaR framework on multiple fact retrieval tasks, showing that it significantly outperforms relevant baselines that use the three-step approach.

\end{abstract}

\section{Introduction}
Knowledge graphs (KGs)~\cite{Freebase, wikidata, dbpedia}, which consist of a set of facts represented in the form of a (head entity, relation, tail entity) triplet, can store a large amount of world knowledge. In natural language applications, language models (LMs)~\cite{bert, gpt3} are commonly used; however, their knowledge internalized in parameters is often incomplete, inaccurate, and outdated. Therefore, several recent works suggest augmenting LMs with facts from KGs, for example, in question answering~\cite{unik-qa, udtqa} and dialogue generation~\cite{spaceefficient, surge}.

\begin{figure}[t]

    \centering
    \includegraphics[width=1.0\columnwidth]{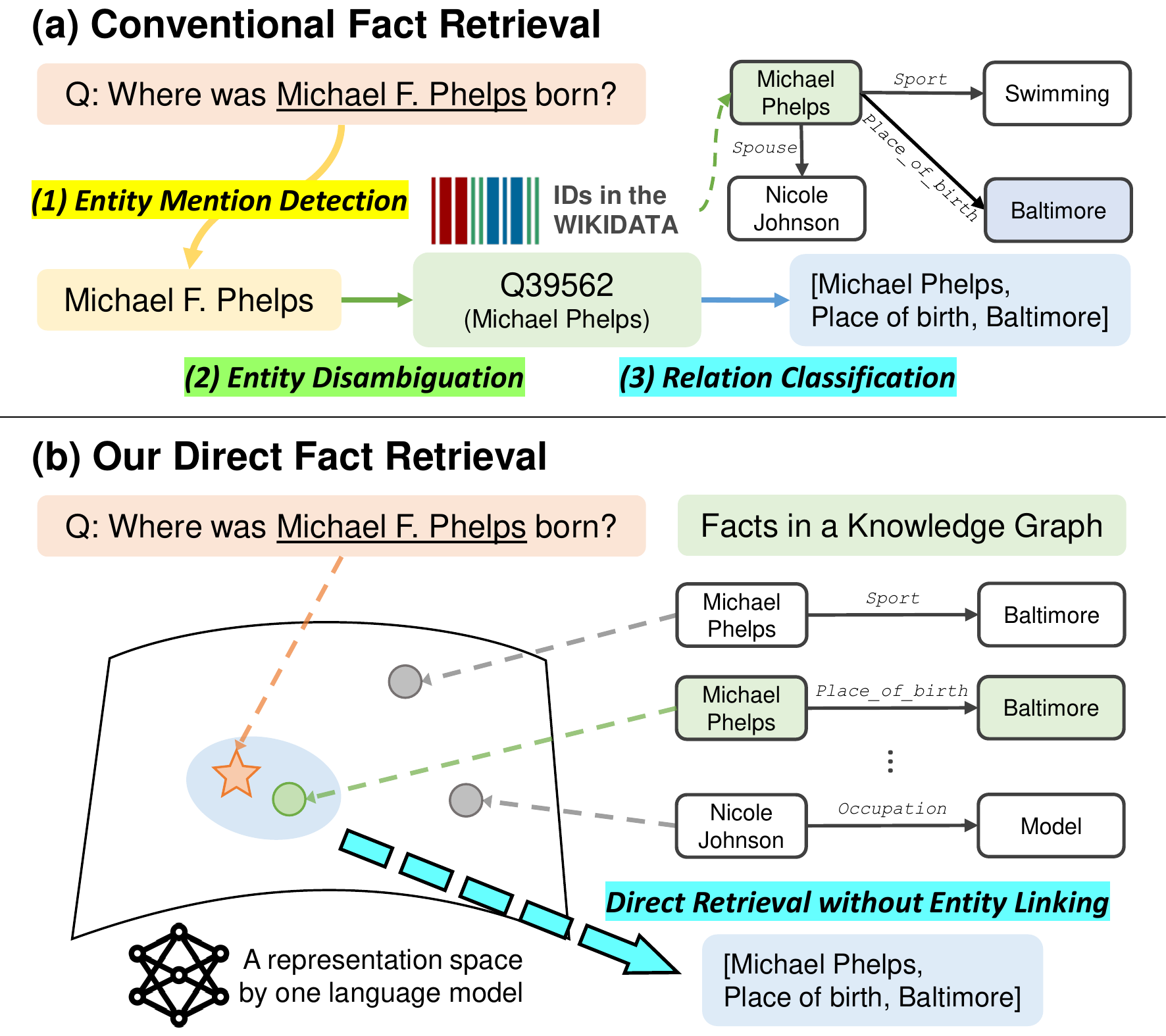}
    \vspace{-0.275in}
    \caption{(a) A conventional fact retrieval from KGs involves three sequential steps: 1) entity mention detection to identify entities in queries; 2) entity disambiguation to match entities in input texts to KGs; 3) relation classification to select relevant relations. (b) Our fact retrieval directly retrieves relevant facts with their representational similarities to input queries.}
    \label{fig:concept}
    \vspace{-0.25in}
    
\end{figure}

However, despite the broad applications of the KGs, the existing mechanism for retrieving facts from them are, in many cases, unnecessarily complex. In particular, to retrieve facts from KGs, existing work~\cite{KBQA/IR/1, KBQA/IR/2, pipeline/retrieve-rerank} relies on three sequential steps, consisting of span detection, entity disambiguation, and relation classification, as illustrated in Figure~\ref{fig:concept}a. For example, given an input text: "Where was Michael Phelps born?", they first detect a span of an entity within the input, which corresponds to "Michael Phelps". Then, they match the entity mention in the input to an entity id in the KG. Those two steps are often called entity linking. Finally, among 91 relations associated with the entity of Michael Phelps, they select one relation relevant to the input, namely "place of birth". 

The aforementioned approach has a couple of drawbacks. First, all three sub-modules in the existing pipeline require module-specific labels in addition to query-triplet pairs for training. However, in real-world, high-quality training data is limited, and annotating them requires significant costs. Second, such a pipeline approach is prone to error propagation across steps~\cite{pipeline/error/1, pipeline/error/2}. For example, if the span detection fails, the subsequent steps, such as relation classification, are likely to make incorrect predictions as well. Third, certain modules, that match entities in queries to KGs or predict relations over KGs, are usually not generalizable to emerging entities and relations and cannot be applied to different KGs. It would be preferable to have a method that does not require KG-specific training and inference.

To tackle these limitations, we propose to directly retrieve the relevant triplets related to a natural language query by computing their similarities over a shared representation space (see Figure~\ref{fig:concept}b). The design of our direct retrieval framework is motivated by a pioneering work of open-domain question answering with documents~\cite{dpr}, which showed the possibility of dense retrieval with simple vector similarities between the question and document embeddings. However, in contrast to the document retrieval scenario where documents have sufficient contexts to embed, it is unclear whether the LM can still effectively embed facts represented in the short triplet form for retrieval. Also, compared to the document retrieval which additionally requires a reader to extract only the relevant piece of knowledge, our fact retriever itself can directly provide the relevant knowledge.

To realize our fact retriever, we train it by maximizing similarities between representations of relevant pairs of input texts and triplets while minimizing irrelevant pairs, where we use LMs for encoding them. We note that this process requires only text-triplet pairs without using extra labels, unlike the conventional pipeline approach for fact retrieval. After training, we index all triplets in the KG with the trained encoder in an offline manner, and, given the input query, we return the nearest triplets over the embedding space. This procedure simplifies the conventional three steps for retrieving facts from KGs into one. To further efficiently search the relevant triplets, we approximate the similarity calculation with vector quantization and hierarchical search based on clustering~\cite{faiss}. We further note that, since we embed triplets using the LM, our retriever can generalize to different KGs without any modification, unlike some conventional retrieval systems that require additional training to learn new KG schema about distinct entities and relations types. We refer to our framework as \textbf{Di}rect \textbf{Fa}ct \textbf{R}etrieval (\textbf{DiFaR}). 

We experimentally demonstrate that our direct retrieval on KGs works well; however, the fact represented in the triplet form has a limited context, since it consists of only two entities and one relation. Also, similarity calculation with the independently represented input text and triplets is arguably simple, and might be less effective. Therefore, to further improve the retriever performance, we additionally use a reranker, whose goal is to calibrate the ranks of retrieved triplets for the input text. In particular, we first retrieve $k$ nearest facts with the direct retriever, and then use another LM which directly measures the similarity by encoding the input text and the triplet simultaneously. Moreover, another objective of the reranker is to filter out irrlevant triplets, which are the most confusing ones in the embedding space of the direct retriever. Therefore, to effectively filter them, we train the reranker to minimize similarities between the input text and the most nearest yet irrelevant triplets. 

We evaluate our DiFaR framework on fact retrieval tasks across two different domains of question answering and dialogue, whose goals are to retrieve relevant triplets in response to the given query. The experimental results show that our DiFaR framework outperforms relevant baselines that use conventional pipeline approaches to retrieve facts on KGs, and also show that our reranking strategy significantly improves retrieval performances. The detailed analyses further support the efficacy of our DiFaR framework, with its great simplicity.

Our contributions in this work are as follows:
\vspace{-0.1in}
\begin{itemize}[itemsep=0.75mm, parsep=1pt]
  \item We present a novel direct fact retrieval (DiFaR) framework from KGs, which leverages only the representational similarities between the query and triplets, simplifying the conventional three steps: entity detection, disambiguation, and relation classification, into one.
  \item We further propose a reranking strategy, to tackle a limitation of little context in facts, for direct knowledge retrieval, which is trained with samples confused by the direct retriever.
  \item We validate our DiFaR on fact retrieval tasks, showing that it significantly outperforms baselines on unsupervised and supervised setups.
\end{itemize}

\section{Background and Related Work}
\label{sec:related_work}

\vspace{-0.05in}
\paragraph{Knowledge Graphs} Knowledge Graphs (KGs) are factual knowledge sources~\cite{Freebase, wikidata}, containing a large number of facts, represented in a symbolic triplet form: (head entity, relation, tail entity). Since some natural language applications require factual knowledge~\cite{kgnlp}, existing literature proposes to use knowledge in KGs, and sometimes along with language models (LMs)~\cite{bert}. To mention a few, in question answering domains, facts in KGs can directly be answers for knowledge graph question answering tasks~\cite{factoidqa/retrieval, KBQA}, but also they are often augmented to LMs to generate knowledge-grounded answers~\cite{ernie, kala}. Similarly, in dialogue generation, some existing work augments LMs with facts from KGs~\cite{spaceefficient, surge}. However, prior to utilizing facts in KGs, fact retrieval -- selection of facts relevant to the input context -- should be done in advance, whose results substantially affect downstream performances. In this work, we propose a conceptually simple yet effective framework for fact retrieval, motivated by information retrieval.

\vspace{-0.05in}
\paragraph{Information Retrieval} The goal of most information retrieval work is to retrieve relevant documents in response to a query (e.g., question). Early work relies on term-based matching algorithms, which count lexical overlaps between the query and documents, such as TF-IDF and BM25~\cite{bm25, bm25/beyond}. However, they are vulnerable to a vocabulary mismatch problem, where semantically relevant documents are lexically different from queries~\cite{vocabmis/1, vocabmis/2}. Due to such the issue, recently proposed work instead uses LMs~\cite{bert, roberta} to encode queries and documents, and uses their representational similarities over a latent space~\cite{dpr, ance, rocketqa}. They suggest their huge successes are due to the effectiveness of LMs in embedding documents. However, they focus on lengthy documents having extensive context, and it is unclear whether LMs can still effectively represent each fact, succinctly represented with two entities and one relation in the triplet form, for its retrieval. In this work, we explore this new direction by formulating the fact retrieval problem as the information retrieval problem done for documents.

\vspace{-0.05in}
\paragraph{Knowledge Retrieval from KGs} Since KGs have a large number of facts, it is important to bring only the relevant piece of knowledge given an input query. To do so, one traditional approach uses neural semantic parsing-based methods~\cite{SP/1, SP/2, SP/3, SP/4} aiming to translate natural language inputs into logical query languages, such as SPARQL\footnote{https://www.w3.org/TR/rdf-sparql-query/} and $\lambda$-DCS~\cite{DCS}, executable over KGs. However, they have limitations in requiring additional labels and an understanding of logical forms of queries. Another approach is to use a pipeline~\cite{pipeline/1, pipeline/2, pipeline/3, pipeline/4, pipeline/retrieve-rerank} consisting of three subtasks: entity span detection, entity disambiguation, and relation classification. However, they similarly require additional labels on training each subcomponent, and this pipeline approach suffers from errors that are propagated from previous steps~\cite{pipeline/error/1, pipeline/error/2}. While recent work~\cite{unik-qa} proposes to retrieve textual triplets from KGs based on their representational similarities to the input text with the information retrieval mechanism, they still rely on entity linking (e.g., span detection and entity disambiguation) first, thus identically having limitations of the pipeline approach. Another recent work~\cite{udtqa} merges a set of facts associated with each entity into a document and performs document-level retrieval. However, the document retrieval itself can be regarded as entity linking, and also the overall pipeline requires an additional reader to extract only the relevant entity in retrieved documents. In contrast to them, we directly retrieve facts from the input query based on their representational similarities, which simplifies the conventional three-step approach including entity linking into one single retrieval step.

\section{DiFaR: Direct Fact Retrieval}

\subsection{Preliminaries}
We formally define a KG and introduce a conventional mechanism for retrieving facts from the KG.

\paragraph{Knowledge Graphs} Let $\mathcal{E}$ be a set of entities and $\mathcal{R}$ be a set of relations. Then, one particular fact is defined as a triplet: $t = (\mathtt{e_h}, \mathtt{r}, \mathtt{e_t}) \in \mathcal{E} \times \mathcal{R} \times \mathcal{E}$, where $\mathtt{e_h}$ and $\mathtt{e_t}$ are head and tail entities, respectively, and $\mathtt{r}$ is a relation between them. Also, a knowledge graph (KG) $\mathcal{G}$ is defined as a set of factual triplets: $\mathcal{G} = \{(\mathtt{e_h}, \mathtt{r}, \mathtt{e_t})\} \subseteq \mathcal{E} \times \mathcal{R} \times \mathcal{E}$. Note that this KG is widely used as a useful knowledge source for many natural language applications, including question answering and dialogue generation~\cite{unik-qa, udtqa, spaceefficient, surge}. However, the conventional mechanism to access facts in KGs is largely complex, which may hinder its broad applications, which we describe in the next paragraph.

\paragraph{Existing Knowledge Graph Retrieval} The input of most natural language tasks is represented as a sequence of tokens: $\vx = [w_1, w_2, \ldots, w_{|\vx|}]$. Suppose that, given the input $\vx$, $t^+$ is a target triplet to retrieve\footnote{For the sake of simplicity, we consider one triplet $t^+$ for each input; the retrieval target can be a set of triplets $\left\{ t^+ \right\}$.}. Then, the objective of the conventional fact retrieval process for the KG $\mathcal{G}$~\cite{pipeline/1, pipeline/retrieve-rerank} is, in many cases, formalized as the following three sequential tasks:
\begin{equation}
\fontsize{10.25pt}{10.25pt}\selectfont
    t^+ = \argmax_{t \in \mathcal{G}} p_\theta(t | \texttt{e}, \vx, \mathcal{G}) p_\phi(\texttt{e} | m, \vx) p_\psi(m | \vx),
\fontsize{10.25pt}{10.25pt}\selectfont
\end{equation}
where $p_\psi(m | \vx)$ is the model for mention detection with $m$ as the detected entity mention within the input $\vx$, $p_\phi(\texttt{e} | m, \vx)$ is the model for entity disambiguation, and $p_\theta(t | \texttt{e}, \vx, \mathcal{G})$ is the model for relation classification, all of which are individually parameterized by $\phi$, $\psi$, and $\theta$, respectively.

However, there is a couple of limitations in such the three-step approaches. First, they are vulnerable to the accumulation of errors, since, for example, if the first two steps consisting of span detection and entity disambiguation are wrong and we are ending up with the incorrect entity irrelevant to the given query, we cannot find the relevant triplet in the final relation prediction stage. Second, due to their decomposed structures, three sub-modules are difficult to train in an end-to-end fashion, while requiring labels for training each sub-module. For example, to train $p_\psi(m | \vx)$ that aims to predict the mention boundary of the entity within the input text, they additionally require annotated pairs of the input text and its entity mentions: $\left\{ (\vx, m) \right\}$. Finally, certain modules are usually limited to predicting entities $\mathcal{E}$ and relations $\mathcal{R}$ specific to the particular KG schema, observed during training. Therefore, they are not directly applicable to unseen entities and relations, but also to different KGs.

\subsection{Direct Knowledge Graph Retrieval}
\label{subsec:retrieval}
To tackle the aforementioned challenges of the existing fact retrieval approaches on KGs, we present the direct knowledge retrieval framework. In particular, our objective is simply formulated with the single sentence encoder model $E_\theta$ without introducing extra variables (e.g., $m$ and $\texttt{e}$), as follows: 
\begin{equation}
    t^+ = \argmax_{t \in \mathcal{G}} f(E_\theta(\vx), E_\theta(t)),
    \label{eq:our_retrieval}
\end{equation}
where $f$ is a non-parametric scoring function that calculates the similarity between the input text representation $E_\theta(\vx)$ and the triplet representation $E_\theta(t)$, for example, by using the dot product. Note that, in Equation~\ref{eq:our_retrieval}, we use the sentence encoder $E_\theta$ to represent the triplet $t$. To do so, we first symbolize the triplet as a sequence of tokens: $t = [w_1, w_2, \ldots, w_{|t|}]$, which is constructed by entity and relation tokens, and the separation token (i.e., a special token, [SEP]) between them. Then, we simply forward the triplet tokens to $E_\theta$ to obtain the triplet representation. While we use the single model for encoding both input queries and triplets, we might alternatively represent them with different encoders, which we leave as future work. 

\paragraph{Training} After formalizing the goal of our direct knowledge retrieval framework in Equation~\ref{eq:our_retrieval}, the next step is to construct the training samples and the optimization objective to train the model (i.e., $E_\theta$). According to Equation~\ref{eq:our_retrieval}, the goal of our model is to minimize distances between the input text and its relevant triplets over an embedding space, while minimizing distances of irrelevant pairs. Therefore, following the existing dense retrieval work for documents~\cite{dpr}, we use a contrastive loss as our objective to generate an effective representation space, formalized as follows:
\begin{equation}
    \min_\theta - \log \frac{ \texttt{exp} ( f(E_\theta(\vx), E_\theta(t^+)) ) }{\sum_{(\vx, t) \in \tau} \texttt{exp} (f(E_\theta(\vx), E_\theta(t))) },
\end{equation}
where $\tau$ contains a set of pairs between the input text and all triplets in the same batch. In other words, $(\vx, t+) \in \tau$ is the positive pair to maximize the similarity, whereas, others are negative pairs to minimize. Also, $\texttt{exp} (\cdot)$ is an exponential function.

\paragraph{Inference} During the inference stage, given the input text $\vx$, the model should return the relevant triplets, whose embeddings are closest to the input text embedding. Note that, since $E_\theta(\vx)$ and $E_\theta(t)$ in Equation~\ref{eq:our_retrieval} are decomposable, to efficiently do that, we represent and index all triplets in an offline manner. Note that, we use the FAISS library~\cite{faiss} for triplet indexing and similarity calculation, since it provides the extremely efficient search logic, also known to be applicable to billions of dense vectors; therefore, suitable for our fact retrieval from KGs. Moreover, to further reduce the search cost, we use the approximated neighborhood search algorithm, namely Hierarchical Navigable Small World Search with Scalar Quantizer. This mechanism not only quantizes the dense vectors to reduce the memory footprint, but also builds the hierarchical graph structures to efficiently find the nearest neighborhoods with few explorations. We term our \textbf{Di}rect \textbf{Fa}ct \textbf{R}etrieval method as \textbf{DiFaR}.

\subsection{Reranking for Accurate Fact Retrieval}
The fact retrieval framework outlined in Section~\ref{subsec:retrieval} simplifies the conventional three subtasks used to access the knowledge into the single retrieval step. However, contrary to the document retrieval case, the fact is represented with the most compact triplet form, which consists of only two entities and one relation. Therefore, it might be suboptimal to rely on the similarity, calculated by the independently represented input text and triplets as in Equation~\ref{eq:our_retrieval}. Also, it is significantly important to find the correct triplet within the small $k$ (e.g., $k=1$) of the top-$k$ retrieved triplets, since, considering the scenario of augmenting LMs with facts, forwarding several triplets to LMs yields huge computational costs. 

To tackle such challenges, we propose to further calibrate the ranks of the retrieved triplets from our DiFaR framework. Specifically, we first obtain the $k$ nearest facts in response to the input query over the embedding space, by using the direct retrieval mechanism defined in Section~\ref{subsec:retrieval}. Then, we use another LM, $E_\phi$, that returns the similarity score of the pair of the input text and the retrieved triplet by encoding them simultaneously, unlike the fact retrieval in Equation~\ref{eq:our_retrieval}. In other words, we first concatenate the token sequences of the input text and the triplet: $[\vx, t]$, where $[\cdot]$ is the concatenation operation, and then forward it to $E_\phi([\vx, t])$. By doing so, the reranking model $E_\phi$ can effectively consider token-level relationships between two inputs (i.e., input queries and triplets), which leads to accurate calibration of the ranks of retrieved triplets from DiFaR, especially for the top-$k$ ranks with small $k$.

For training, similar to the objective of DiFaR defined in Section~\ref{subsec:retrieval}, we aim to maximize the similarities of positive pairs: $\left\{ (\vx, t^+) \right\}$, while minimizing the similarities of irrelevant pairs: $\left\{ (\vx, t) \right\} \setminus \left\{ (\vx, t^+) \right\}$. To do so, we use a binary cross-entropy loss. However, contrary to the previous negative sampling strategy defined in Section~\ref{subsec:retrieval} where we randomly sample the negative pairs, in this reranker training, we additionally manipulate them by using the initial retrieval results from our DiFaR. The intuition here is that the irrelevant triplets, included in the $k$ nearest neighbors to the input query, are the most confusing examples, which are yet not filtered by the DiFaR model. Hereat, the goal of the reranking strategy is to further filter them by refining the ranks of the $k$ retrieved triplets; therefore, to achieve this goal, we include them as the negative samples during reranker training. Formally, let $\tilde{\tau} = \left\{ (\vx, \tilde{t}) \right\}$ is a set of pairs of the input query $\vx$ and its $k$ nearest facts retrieved from DiFaR. Then, the negative samples for the reranker are defined by excluding the positive pairs, formalized as follows: $\tilde{\tau} \setminus \left\{ (\vx, t^+) \right\}$. Note that constructing the negative samples with retrieval at every training iteration is costly; therefore, we create them at intervals of several epochs (e.g., ten), but also we use only a subset of triplets in KGs during retrieval. Our framework with the reranking strategy is referred to as \textbf{Di}rect \textbf{Fa}ct \textbf{R}etrieval with \textbf{R}eranking (\textbf{DiFaR}\boldmath{$^2$}).

\section{Experimental Setups}

We explain datasets, models, metrics, and implementations. For additional details, see Appendix~\ref{sec:setups}.

\subsection{Datasets}

We validate our \textbf{Di}rect \textbf{Fa}ct \textbf{R}etrieval (\textbf{DiFaR}) on fact retrieval tasks, whose goal is to retrieve relevant triplets over KGs given the query. We use four datasets on question answering and dialogue tasks. 

\paragraph{Question Answering} The goal of KG-based question answering (QA) tasks is to predict factual triplets in response to the given question, where predicted triplets are direct answers. For this task, we use three datasets, namely SimpleQuestions~\cite{SimpleQuestions}, WebQuestionsSP (WebQSP)~\cite{WebQuestions, WebQSP}, and Mintaka~\cite{mintaka}. Note that SimpleQuestions and WebQSP are designed with the Freebase KG~\cite{Freebase}, ad Mintaka is designed with the Wikidata KG~\cite{wikidata}.

\begin{table*}[t!]
\caption{\textbf{Main results on the question answering domain} for SimpleQuestions, WebQSP, and Mintaka datasets. We emphasize the best scores in bold, except for the incomparable model: Retrieval with Gold Entities, which uses labeled entities in inputs.}
\vspace{-0.1in}
\label{tab:main:qa}
\small
\centering
\resizebox{\textwidth}{!}{
\renewcommand{\arraystretch}{1.0}
\begin{tabular}{llcccccccccccc}
\toprule

& & \multicolumn{3}{c}{\bf SimpleQuestions} & \multicolumn{3}{c}{\bf WebQSP} & \multicolumn{3}{c}{\bf Mintaka} \\
\cmidrule(l{2pt}r{2pt}){3-5} \cmidrule(l{2pt}r{2pt}){6-8} \cmidrule(l{2pt}r{2pt}){9-11}
\textbf{Types} & \textbf{Methods} & MRR & Hits@1 & Hits@10 & MRR & Hits@1 & Hits@10 & MRR & Hits@1 & Hits@10 \\

\midrule
\midrule

\multirowcell{8}[-0.5ex][l]{\textbf{Unsupervised}} 

& Retrieval with Gold Entities
& 0.7213 & 0.5991 & 0.9486 & 0.5324 & 0.4355 & 0.7402 & 0.1626 & 0.0978 & 0.2969 \\

\noalign{\vskip 0.25ex}\cdashline{2-14}\noalign{\vskip 0.75ex}

& Retrieval with spaCy
& 0.3454 & 0.2917 & 0.4437 & 0.3530 & 0.2856 & 0.4863 & 0.0914 & 0.0585 & 0.1622 \\

& Retrieval with GENRE
& 0.1662 & 0.1350 & 0.2234 & 0.3099 & 0.2498 & 0.4363 & 0.0935 & 0.0640 & 0.1540 \\

& Retrieval with BLINK
& 0.5142 & 0.4276 & 0.6766 & 0.4853 & 0.3938 & 0.6694 & 0.1350 & 0.0850 & 0.2430 \\

& Retrieval with ReFinED
& 0.4841 & 0.4047 & 0.6283 & 0.5008 & 0.4055 & 0.6953 & 0.1312 & 0.0831 & 0.2325 \\

& Factoid QA by Retrieval 
& 0.7835 & 0.6953 & 0.9304 & 0.3933 & 0.3089 & 0.5470 & 0.1350 & 0.0836 & 0.2344 \\ 

\noalign{\vskip 0.25ex}\cdashline{2-14}\noalign{\vskip 0.75ex}

& \textbf{DiFaR (Ours)}
& 0.7070 & 0.5872 & 0.9259 & 0.5196 & 0.4130 & 0.7352 & 0.1590 & 0.0895 & 0.3043  \\

& \textbf{DiFaR\boldmath{$^2$} (Ours)}
& \textbf{0.8361} & \textbf{0.7629} & \textbf{0.9470} & \textbf{0.5441} & \textbf{0.4321} & \textbf{0.7602} & \textbf{0.2077} & \textbf{0.1348} & \textbf{0.3595} \\

\midrule

\multirowcell{8}[-0.5ex][l]{\textbf{Supervised}} 

& Retrieval with Gold Entities 
& 0.8007 & 0.7094 & 0.9477 & 0.6048 & 0.5079 & 0.7794 & 0.2705 & 0.1987 & 0.4070 \\

\noalign{\vskip 0.25ex}\cdashline{2-14}\noalign{\vskip 0.75ex}

& Retrieval with spaCy
& 0.3789 & 0.3380 & 0.4453 & 0.3963 & 0.3272 & 0.5162 & 0.1367 & 0.1019 & 0.2019 \\

& Retrieval with GENRE
& 0.1921 & 0.1718 & 0.2255 & 0.3617 & 0.3014 & 0.4696 & 0.1346 & 0.1005 & 0.1964 \\

& Retrieval with BLINK
& 0.5679 & 0.5008 & 0.6766 & 0.5483 & 0.4571 & 0.7052 & 0.2075 & 0.1530 & 0.3157 \\

& Retrieval with ReFinED
& 0.5349 & 0.4765 & 0.6279 & 0.5707 & 0.4754 & 0.7377 & 0.2106 & 0.1562 & 0.3166 \\

& Factoid QA by Retrieval 
& 0.8590 & 0.8051 & 0.9293 & 0.5253 & 0.4546 & 0.6486 & 0.1548 & 0.1179 & 0.2179 \\

\noalign{\vskip 0.25ex}\cdashline{2-14}\noalign{\vskip 0.75ex}

& \textbf{DiFaR (Ours)}
& 0.7904 & 0.6986 & 0.9382 & 0.6102 & 0.5071 & 0.7927 & 0.3049 & 0.2138 & 0.4856 \\

& \textbf{DiFaR\boldmath{$^2$} (Ours)}
& \textbf{0.8992} & \textbf{0.8583} & \textbf{0.9576} & \textbf{0.7189} & \textbf{0.6528} & \textbf{0.8385} & \textbf{0.4189} & \textbf{0.3367} & \textbf{0.5847} \\

\bottomrule

\end{tabular}
}
\vspace{-0.15in}
\end{table*}

\paragraph{Dialogue} In addition to QA, we evaluate our DiFaR on KG-based dialogue generation, whose one subtask is to retrieve relevant triplets on the KG that provides factual knowledge to respond to a user's conversation query. We use the OpenDialKG data~\cite{opendialkg}, designed with Freebase.

\vspace{-0.02in}
\paragraph{Knowledge Graphs} Following~\citet{fbtowiki} and~\citet{rigel/e2e}, we use the Wikidata KG~\cite{wikidata} for our experiments on QA, and use their dataset processing settings. For OpenDialKG, we use Freebase.

\vspace{-0.02in}
\subsection{Baselines and Our Models}
\vspace{-0.02in}

We compare our DiFaR framework against other relevant baselines that involve subtasks, such as entity detection, disambiguation, and relation prediction. Note that most existing fact retrieval work either uses labeled entities in queries, or uses additional labels for training subcomponents; therefore, they are not comparable to DiFAR that uses only pairs of input texts and relevant triplets. For evaluations, we include models categorized as follows:

\paragraph{Retrieval with Entity Linking:} It predicts relations over candidate triplets associated with identified entities by the entity linking methods, namely \textbf{spaCy}~\cite{spacy}, \textbf{GENRE}~\cite{genre}, \textbf{BLINK}~\cite{blink, blink/elq}, and \textbf{ReFinED}~\cite{refined} for Wikidata; \textbf{GrailQA}~\cite{grail} for Freebase. 

\paragraph{Factoid QA by Retrieval:} It retrieves entities and relations independently based on their similarities with the input query~\cite{factoidqa/retrieval}. 

\paragraph{Our Models:} Our \textbf{Di}rect \textbf{K}nowl\textbf{e}dge \textbf{R}etrieval (\textbf{DiFaR}) directly retrieves the nearest triplets to the input text on the latent space. \textbf{DiFaR with Reranking (DiFaR\boldmath{$^2$})} is also ours, which includes a reranker to calibrate retrieved results. 

\paragraph{Retrieval with Gold Entities:} It uses labeled entities in inputs and retrieves triplets based on their associated triplets. It is incomparable to others.

\subsection{Evaluation Metrics}
We measure the retrieval performances of models with standard ranking metrics, which are calculated by ranks of correctly retrieved triplets. In particular, we use \textbf{Hits@K} which measures whether retrieved Top-K triplets include a correct answer or not, and Mean Reciprocal Rank (\textbf{MRR}) which measures the rank of the first correct triplet for each input text and then computes the average of reciprocal ranks of all results. Following exiting document retrieval work~\cite{ance, DAR}, we consider top-1000 retrieved triplets when calculating MRR, since considering ranks of all triplets in KGs are computationally prohibitive.

\subsection{Implementation Details}
\label{subsec:details}
We use a distilbert\footnote{\tiny https://huggingface.co/sentence-transformers/msmarco-distilbert-base-v3} as a retriever for all models, and a lightweight MiniLM model\footnote{\tiny https://huggingface.co/cross-encoder/ms-marco-MiniLM-L-6-v2} as a reranker, both of which are pre-trained with the MSMARCO dataset~\cite{msmarco}. During reranking, we sample top-100 triplets retrieved from DiFaR. We use off-the-shelf models for unsupervised settings, and further train them for supervised settings. 

\begin{table}[t!]
\caption{\textbf{Main results on the dialogue domain} for the OpenDialKG dataset. We emphasize the best scores in bold except for Retrieval with Gold Entities, which uses labeled entities.}
\vspace{-0.1in}
\label{tab:main:dial}
\small
\centering
\resizebox{0.475\textwidth}{!}{
\renewcommand{\arraystretch}{1.1}
\begin{tabular}{llccc}
\toprule

& & \multicolumn{3}{c}{\bf OpenDialKG} \\
\cmidrule(l{2pt}r{2pt}){3-5}
\textbf{Types} & \textbf{Methods} & MRR & Hits@1 & Hits@10 \\

\midrule
\midrule

\multirowcell{5}[-0.5ex][l]{\textbf{Unsupervised}} 

& Retrieval with Gold Entities
& 0.2511 & 0.1560 & 0.4683 \\

\noalign{\vskip 0.25ex}\cdashline{2-5}\noalign{\vskip 0.75ex}

& Retrieval with GrailQA
& 0.2051 & 0.1271 & 0.3745 \\ 

& Factoid QA by Retrieval 
& 0.1977 & 0.0892 & 0.4231 \\ 

\noalign{\vskip 0.25ex}\cdashline{2-5}\noalign{\vskip 0.75ex}

& \textbf{DiFaR (Ours)}
& 0.2396 & 0.1395 & 0.4424 \\

& \textbf{DiFaR\boldmath{$^2$} (Ours)}
& \textbf{0.2637} & \textbf{0.1603} & \textbf{0.4744} \\

\midrule

\multirowcell{5}[-0.5ex][l]{\textbf{Supervised}} 

& Retrieval with Gold Entities
& 0.2750 & 0.1495 & 0.5745 \\

\noalign{\vskip 0.25ex}\cdashline{2-5}\noalign{\vskip 0.75ex}

& Retrieval with GrailQA
& 0.2217 & 0.1198 & 0.4436 \\ 

& Factoid QA by Retrieval 
& 0.2042 & 0.1266 & 0.3587 \\ 

\noalign{\vskip 0.25ex}\cdashline{2-5}\noalign{\vskip 0.75ex}

& \textbf{DiFaR (Ours)}
& 0.2755 & 0.1405 & 0.5547 \\

& \textbf{DiFaR\boldmath{$^2$} (Ours)}
& \textbf{0.4784} & \textbf{0.3535} & \textbf{0.7380} \\

\bottomrule

\end{tabular}
}
\vspace{-0.15in}
\end{table}

\section{Experimental Results and Analyses}
\label{sec:exp}

\paragraph{Main Results}
We first conduct experiments on question answering domains, and report the results in Table~\ref{tab:main:qa}. As shown in Table~\ref{tab:main:qa}, our DiFaR with Reranking (DiFaR$^2$) framework significantly outperforms all baselines on all datasets across both unsupervised and supervised experimental settings with large margins. Also, we further experiment on dialogue domain, and report results in Table~\ref{tab:main:dial}. As shown in Table~\ref{tab:main:dial}, similar to the results on QA domains, our DiFaR$^2$ framework outperforms the relevant baselines substantially. These results on two different domains demonstrate that our DiFaR$^2$ framework is highly effective in fact retrieval tasks.

To see the performance gains from our reranking strategy, we compare the performances between our model variants: DiFaR and DiFaR$^2$. As shown in Table~\ref{tab:main:qa} and Table~\ref{tab:main:dial}, compared to DiFaR, DiFaR$^2$ including the reranker brings huge performance improvements, especially on the challenging datasets: Mintaka and OpenDialKG. However, we consistently observe that our DiFaR itself can also show superior performances against all baselines except for the model of Factoid QA by Retrieval on the SimpleQuestions dataset. The inferior performance of our DiFaR on this SimpleQuestions dataset is because, its samples are automatically constructed from facts in KGs; therefore, it is extremely simple to extract entities and predict relations in response to the input query. On the other hand, our DiFaR framework sometimes outperforms the incomparable model: Retrieval with Gold Entities, which uses the labeled entities in the input queries. This is because this model is restricted to retrieve the facts that should be associated with entities in input queries; meanwhile, our DiFaR is not limited to query entities thanks to the direct retrieval scheme.

\begin{table}[t!]
\caption{\textbf{Zero-shot transfer learning results.} We use models trained on the WebQSP dataset with the Wikidata KG not only for SimpleQuestions and Mintaka datasets with the same KG, but also for the WebQSP dataset with the different Freebase KG. We use MRR as a metric, and N/A denotes not available.}
\vspace{-0.1in}
\label{tab:transfer}
\small
\centering
\resizebox{0.475\textwidth}{!}{
\renewcommand{\arraystretch}{1.1}
\begin{tabular}{lcccc}
\toprule

& \multicolumn{2}{c}{\bf Wikidata} & \multicolumn{2}{c}{\bf Freebase} \\
\cmidrule(l{2pt}r{2pt}){2-3} \cmidrule(l{2pt}r{2pt}){4-4}
\textbf{Methods} & SimpleQuestions & Mintaka & WebQSP \\

\midrule
\midrule

Retrieval with Gold Entities
& 0.7994 & 0.1950 & 0.6000 \\

\noalign{\vskip 0.25ex}\cdashline{1-4}\noalign{\vskip 0.75ex}

Retrieval with BLINK
& 0.5704 & 0.1617 & N/A \\ 

Retrieval with ReFinED
& 0.5389 & 0.1591 & N/A \\ 

Factoid QA by Retrieval 
& 0.8014 & 0.1431 & 0.4239 \\ 

\noalign{\vskip 0.25ex}\cdashline{1-4}\noalign{\vskip 0.75ex}

\textbf{DiFaR (Ours)}
& 0.7812 & 0.2063 & 0.5913 \\

\textbf{DiFaR\boldmath{$^2$} (Ours)}
& \textbf{0.8244} & \textbf{0.2769} & \textbf{0.6324} \\

\bottomrule

\end{tabular}
}
\vspace{-0.05in}
\end{table}

\begin{figure}[t!]
    \centering
    \includegraphics[width=0.99\columnwidth]{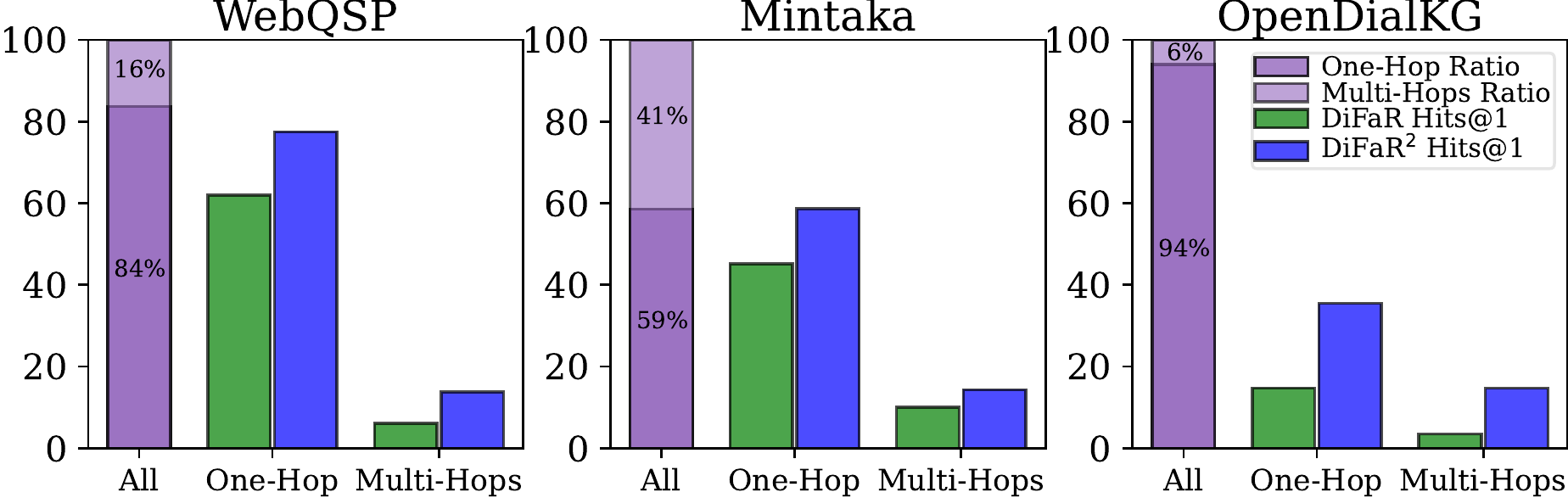}
    \vspace{-0.275in}
    \caption{\textbf{Breakdown results by single and multi-hops.} We report ratios of single and multi-hops samples on the left side of each subfigure, and Hits@1 of DiFaR and DiFaR$^2$ across single and multi-hops on the middle and right. We exclude the SimpleQuestions dataset that consists of single-hop questions.}
    \label{fig:multihop}
    \vspace{-0.15in}
\end{figure}

\paragraph{Analyses on Zero-Shot Generalization} Our DiFaR can be generalizable to different datasets with the same KG, but also to ones with other KGs without any modifications. This is because it retrieves triplets based on their text-level similarities to input queries and does not leverage particular schema of entities and relations, unlike the existing entity linking methods. To demonstrate them, we perform experiments on zero-shot transfer learning, where we use the model, trained on the WebQSP dataset with the Wikidata KG, to different datasets with the same KG and also to ones with the different Freebase KG. As shown in Table~\ref{tab:transfer}, our DiFaR frameworks are effectively generalizable to different datasets and KGs; meanwhile, the pipeline approaches involving entity linking are not generalizable to different KGs, and inferior to ours.

\paragraph{Analyses on Single- and Multi-Hops}
To see whether our DiFaR frameworks can also perform challenging multi-hop retrieval that requires selecting triplets not directly associated with entities in input queries, we breakdown the performances by single- and multi-hop type queries. As shown in Figure~\ref{fig:multihop}, our DiFaR can directly retrieve relevant triplets regardless of whether they are associated with entities in input queries (single-hop) or not (multi-hop), since it does not rely on entities in queries for fact retrieval. Also, we observe that our reranking strategy brings huge performance gains, especially on multi-hop type queries. However, due to the intrinsic complexity of multi-hop retrieval, its performances are relatively lower than performances in single-hop cases. Therefore, despite the fact that the majority of queries are answerable with single-hop retrieval and that our DiFaR can handle multi-hop queries, it is valuable to further extend the model for multi-hop, which we leave as future work. We also provide examples of facts retrieved by our DiFaR framework in Table~\ref{tab:example}. As shown in Table~\ref{tab:example}, since LMs, that is used for encoding both the question and the triplets for retrieval, might learn background knowledge about them during pre-trainnig, our DiFaR framework can directly retrieve relevant triplets even for complex questions. For instance, in the first example of Table~\ref{tab:example}, the LM already knows who was the us president in 1963, and directly retrieves whose religion. Additionally, we provide more retrieval examples of our DiFaR framework in Appendix~\ref{appendix:examples} with Table~\ref{tab:examples_additional} for both single- and multi-hop questions.

\begin{table}[t!]
\centering
\Large
\caption{\small \textbf{Retrieval examples for complex questions}, on the challenging Mintaka dataset. We highlight the related phrases across the question and the triplet in yellow and green colors.}
\vspace{-0.1in}
\resizebox{0.48\textwidth}{!}{
\renewcommand{\arraystretch}{0.8}
\begin{tabular}{lll}

\toprule
\multicolumn{3}{p{\textwidth}}{\textbf{Question}: What \hlyellow{religion} was \hlgreen{the us president in 1963}?} \\

\multicolumn{3}{p{\textwidth}}{\textbf{Retrieved Triplet}: (\hlgreen{Robert F. Kennedy}, religion, \hlyellow{Catholicism})} \\

\multicolumn{3}{p{\textwidth}}{\textbf{Answer}: Catholicism} \\

\midrule

\multicolumn{3}{p{\textwidth}}{\textbf{Question}: Who commanded \hlyellow{the allied invasion of western Europe at Normandy} and was an \hlgreen{American president}?} \\

\multicolumn{3}{p{\textwidth}}{\textbf{Retrieved Triplet}: (\hlyellow{Normandy landings}, participant, \hlgreen{Dwight D. Eisenhower})} \\

\multicolumn{3}{p{\textwidth}}{\textbf{Answer}: Dwight D. Eisenhower} \\

\midrule

\multicolumn{3}{p{\textwidth}}{\textbf{Question}: Which \hlyellow{former Chicago Bull shooting guard} was also selected to play on \hlgreen{the 1992 US basketball team}?} \\

\multicolumn{3}{p{\textwidth}}{\textbf{Retrieved Triplet}: (\hlyellow{1992 US men's basketball team}, has part, \hlgreen{Michael Jordan})} \\

\multicolumn{3}{p{\textwidth}}{\textbf{Answer}: Michael Jordan} \\

\bottomrule
\end{tabular}
}
\label{tab:example}
\vspace{-0.05in}
\end{table}

\begin{figure}[t!]
    \centering
    \includegraphics[width=1\columnwidth]{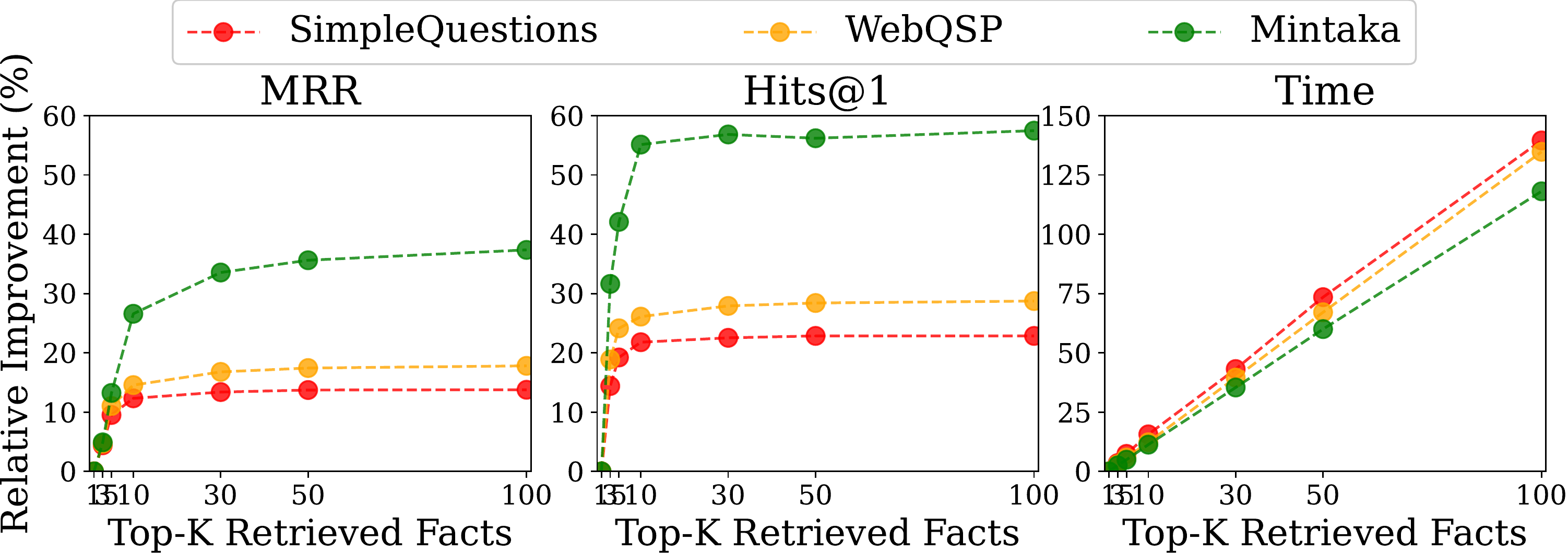}
    \vspace{-0.28in}
    \caption{\textbf{Performances and efficiencies of DiFaR\boldmath{$^2$} with varying K}, where we change the number of Top-K retrieved triplets when leveraging the reranking strategy. We report results with the relative improvement (\%) to our DiFaR without reranking. We report the time with average over 30 runs.}
    \label{fig:efficiency}
    \vspace{-0.15in}
\end{figure}

\paragraph{Analyses on Reranking with Varying K} While we show huge performance improvements with our reranking strategy in Table~\ref{tab:main:qa} and Table~\ref{tab:main:dial}, its performances and efficiencies depend on the number of retrieved Top-K triplets. Therefore, to further analyze it, we vary the number of K, and report the performances and efficiencies in Figure~\ref{fig:efficiency}. As shown in Figure~\ref{fig:efficiency}, the performances are rapidly increasing until Top-10 and saturated after it. Also, the time for reranking is linearly increasing when we increase the K values, and, in Top-10, the reranking mechanism takes only less than 20\% time required for the initial retrieval. These results suggest that it might be beneficial to set the K value as around 10.

\paragraph{Sensitivity Analyses on Architectures} To see different architectures of retrievers and rerankers make how many differences in performances, we perform sensitivity analyses by varying their backbones. We use available models in the huggingface model library\footnote{https://huggingface.co/models}. As shown in Table~\ref{tab:sensitivity}, we observe that the pre-trained backbones by the MSMARCO dataset~\cite{msmarco} show superior performances compared to using the naive backbones, namely DistilBERT and MiniLM, on both retrievers and rerankers. Also, performance differences between models with the same pre-trained dataset (e.g., MSMARCO-TAS-B and MSMARCO-Distil) are marginal. These two results suggest that the knowledge required for document retrieval is also beneficial to fact retrieval, and that DiFaR frameworks are robust across different backbones.

\begin{table}[t!]
\caption{\textbf{Sensitivity analyses on architectures}, where we change the backbones of retriever and reranker in our DiFaR$^2$. MSMARCO in the model name indicates it is pre-trained by the MSMARCO dataset, and we report results on the WebQSP.}
\vspace{-0.1in}
\label{tab:sensitivity}
\small
\centering
\resizebox{0.48\textwidth}{!}{
\renewcommand{\arraystretch}{1.1}
\begin{tabular}{llccc}
\toprule

\textbf{Types} & \textbf{Models} & MRR & Hits@1 & Hits@10 \\

\midrule
\midrule

\multirowcell{3}[-0.5ex][l]{\textbf{Retriever}} 
& DistilBERT
& 0.5983 & 0.4963 & 0.7810 \\

& MSMARCO-TAS-B
& 0.6051 & 0.4963 & 0.7844 \\

& MSMARCO-Distil
& 0.6102 & 0.5071 & 0.7927 \\

\midrule

\multirowcell{3}[-0.5ex][l]{\textbf{Reranker}} 
& MiniLM
& 0.6675 & 0.5945 & 0.7927 \\

& MSMARCO-TinyBERT
& 0.7068 & 0.6420 & 0.8177 \\

& MSMARCO-MiniLM
& 0.7189 & 0.6528 & 0.8385 \\

\bottomrule

\end{tabular}
}
\vspace{-0.05in}
\end{table}
\begin{figure}[t!]
    \centering
    \includegraphics[width=1\columnwidth]{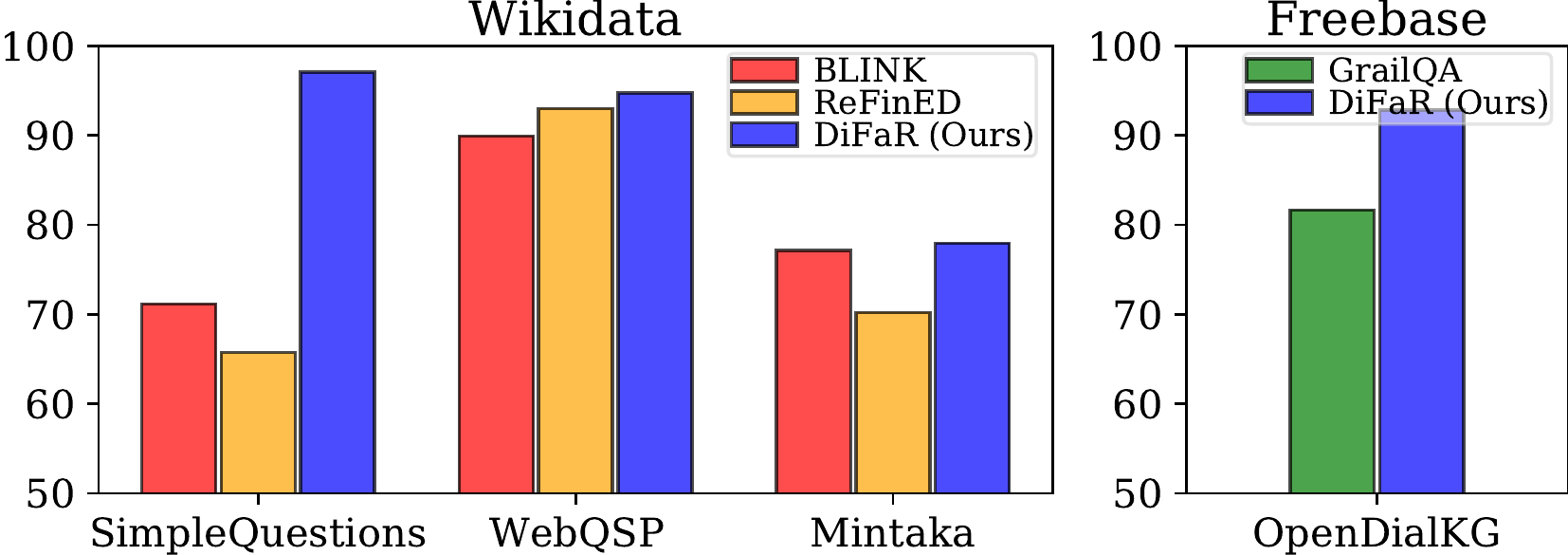}
    \vspace{-0.27in}
    \caption{\textbf{Entity linking results,} where we measure the performances on benchmark datasets with Wikidata and Freebase KGs. Note that entity mentions of the SimpleQuestions dataset are not available; therefore, we cannot fine-tune existing entity linkers, which additionally require mention labels, unlike ours.}
    \label{fig:linking}
    \vspace{-0.175in}
\end{figure}

\paragraph{Analyses on Entity Linking} While our DiFaR framework is not explicitly trained to predict entity mentions in the input query and their ids in the KG, during the training of our DiFaR, it might learn the knowledge on matching the input text to its entities. To demonstrate it, we measure entity linking performances by checking whether the retrieved triplets contain the labeled entities in the input query. As shown in Figure~\ref{fig:linking}, our DiFaR surprisingly outperforms entity linking models. This might be because there are no accumulation of errors in entity linking steps, which are previously done with mention detection and entity disambiguation, thanks to direct retrieval with end-to-end learning; but also the fact in the triplet form has more beneficial information to retrieve contrary to the entity retrieval.

\section{Conclusion}

In this work, we focused on the limitations of the conventional fact retrieval pipeline, usually consisting of entity mention detection, entity disambiguation and relation classification, which not only requires additional labels for training each subcomponent but also is vulnerable to the error propagation across submodules. To this end, we proposed the extremely simple Direct Fact Retrieval (DiFaR) framework. During training, it requires only pairs of input texts and relevant triplets, while, in inference, it directly retrieves relevant triplets based on their representational similarities to the given query. Further, to calibrate the ranks of retrieved triplets, we proposed to use a reranker. We demonstrated that our DiFaR outperforms existing fact retrieval baselines despite its great simplicity, but also ours with the reranking strategy significantly improves the performances; for the first time, we revealed that fact retrieval can be easily yet effectively done. We believe our work paves new avenues for fact retrieval, which leads to various follow-up work.

\section*{Limitations}
\label{sec:limitation}
In this section, we faithfully discuss the current limitations and potential avenues for future research.

First of all, while one advantage of our Direct Fact Retrieval (DiFaR) is its simplicity, this model architecture is arguably simple and might be less effective in handling very complex queries~\cite{mintaka}. For example, as shown in Figure~\ref{fig:multihop}, even though our DiFaR framework can handle the input queries demanding multi-hop retrieval, the performances on such queries are far from perfect. Therefore, future work may improve DiFaR by including more advanced techniques, for example, further traversing over the KG based on the retrieved facts from our DiFaR. Also, while we use only the text-based similarities between queries and triplets with LMs, it is interesting to model triplets over KGs based on their graph structures and blend their representations with representations from LMs to generate more effective search space.  

Also, we focus on retrieval datasets in English. Here we would like to note that, in fact retrieval, most datasets are annotated in English, and, based on this, most existing work evaluates model performances on English samples. However, handling samples in various languages is an important yet challenging problem, and, as future work, one may extend our DiFaR to multilingual settings.

\section*{Ethics Statement}
For an input query, our Direct Fact Retrieval (DiFaR) framework enables the direct retrieval of the factual knowledge from knowledge graphs (KGs), simplifying the conventional pipeline approach consisting of entity detection, entity disambiguation, and relation classification. However, the performance of our DiFaR framework is still not perfect, and it may retrieve incorrect triplets in response to given queries. Therefore, for the high-risk domains, such as biomedicine, our DiFaR should be carefully used, and it might be required to analyze retrieved facts before making the critical decision. 

\section*{Acknowledgements}
We thank the members of the End-to-End Reasoning team of Alexa AI at Amazon and the anonymous reviewers for their constructive and insightful comments. Any opinions, findings, and conclusions expressed in this material are those of the authors and do not necessarily reflect the previous and current funding agencies of the authors. The part of Jinheon Baek's graduate study and, accordingly, this work was supported by the Institute of Information \& communications Technology Planning \& Evaluation (IITP) grant funded by the Korea government (MSIT) (No.2019-0-00075, Artificial Intelligence Graduate School Program (KAIST) and No.2021-0-02068, Artificial Intelligence Innovation Hub), and the Engineering Research Center Program through the National Research Foundation of Korea (NRF) funded by the Korea Government (MSIT) (NRF-2018R1A5A1059921). 

\bibliography{custom}
\bibliographystyle{acl_natbib}

\clearpage

\appendix

\section{Additional Experimental Setups}
\label{sec:setups}
Here we provide additional experimental setups. 

\subsection{Datasets}

\paragraph{Question Answering} In KG-based question answering datasets, there exist pairs of questions and their relevant triplets, and we use them for training and evaluating models. We use the following three datasets: SimpleQuestions~\cite{SimpleQuestions}, WebQuestionsSP (WebQSP)~\cite{WebQuestions, WebQSP}, and Mintaka~\cite{mintaka}, and here we describe them in details. First of all, the SimpleQuestions dataset is designed with the Freebase KG~\cite{Freebase}, which consists of 19,481, 2,821, and 5,622 samples on training, validation, and test sets. Similarly, the WebQSP dataset, which is a refined from the WebQuestions dataset by filtering out samples with invalid annotations, is annotated with the Freebase KG, consisting of 2,612 and 1,375 samples on training and test sets, and we further sample 20\% of training samples for validation. Lastly, the Mintaka dataset is recently designed for complex question answering, which is collected from crowdsourcing and annotated with the Wikidata KG~\cite{wikidata}. Among eight different languages, we use questions in English, which consist of 14,000, 2,000, and 4,000 samples for training, validation, and test sets, respectively. 

\paragraph{Dialogue} Similar to the KG-based question answering datasets, the dataset on KG-based dialogue generation domain has pairs of the input query and its relevant triplets, where the input query consists of the user's utterance and dialogue history, and the annotated triplets are the useful knowledge source to answer the query. For this dialogue domain, we use the OpenDialKG dataset~\cite{opendialkg}, which is collected with two parallel corpora of open-ended dialogues and a Freebase KG. We randomly split the dataset into training, validation, and test sets with ratios of 70\%, 15\%, and 15\%, respectively, and preprocess it following~\citet{surge}, which results in 31,145, 6,722, and 6,711 samples on training, validation, and test sets.

\paragraph{Knowledge Graphs} Following experimental setups of ~\citet{fbtowiki} and~\citet{rigel/e2e}, we use the Wikidata KG~\cite{wikidata} for our experiments on question answering, since the Freebase KG~\cite{Freebase} is outdated, and the recently proposed entity linking models are implemented with the Wikidata, i.e., they are not suitable for the Freebase KG. Specifically, to use the Wikidata KG for datasets designed with the Freebase KG (e.g., SimpleQuestions and WebQSP), we use available mappings from the Freebase KG to the Wikidata KG~\cite{fbtowiki}. Also, we use the wikidata dump of Mar. 07, 2022, and follow the dataset preprocessing setting from~\citet{rigel/e2e}. For the OpenDialKG dataset, since it does not provide the Freebase entity ids, we cannot map them to the Wikidata entity ids using the available entity mappings. Therefore, for this dataset, we use original samples annotated with the Freebase KG. 

\subsection{Baselines and Our Model}
In this subsection, we provide the detailed explanation of models that we use for baselines. Note that entity linking models are further coupled with the relation classification module to predict triplets based on identified entities in input queries. We begin with the explanations of entity linkers.

\paragraph{spaCy} This model~\cite{spacy} sequentially predicts spans and KG ids of entities based on the named entity recognition and entity disambiguation modules. We use the spaCy v3.4\footnote{https://spacy.io/api/entitylinker}.

\paragraph{GENRE} This model~\cite{genre} first predicts the entity spans and then generates the unique entities in an autoregressive manner. Note that this model is trained for long texts; therefore, it may not be suitable for handling short queries.

\paragraph{BLINK} This model~\cite{blink} retrieves the entities based on their representational similarities with the input queries, and, before that, entity mentions in the input should be provided. We use a model further tuned for questions~\cite{blink/elq}.

\paragraph{ReFinED} This model~\cite{refined} performs the entity mention detection and the entity disambiguation in a single forward pass. We use a model further fine-tuned for questions. 

\paragraph{GrailQA} Unlike the above entity linkers that are trained for the Wikidata KG, this model~\cite{grail} is trained to predict entities in the Freebase KG. This model performs the entity detection and the disambiguation sequentially, which is similar to the entity linking mechanism of spaCy.

\paragraph{Factoid QA by Retrieval} This model is a baseline~\cite{factoidqa/retrieval} that individually retrieves the entities and relations based on their embedding-level similarities to input queries. Then, it merges the retrieved entities and relations with the KG-specific schema to construct the triplets.

\paragraph{DiFaR} This is our fact retrieval framework that directly retrieves the facts on KGs based on their representational similarities to the input queries.

\paragraph{DiFaR\boldmath{$^2$}} This is our fact retrieval framework with the proposed reranking strategy, where we further calibrate the retrieved results from DiFaR.

\paragraph{Retrieval with Gold Entities} This is an incomparble model to others, which uses labeled entities in input queries to predict relations based on them.

\subsection{Implementation Details}
\label{subsec:detail}
In this subsection, we provide additional implementation details that are not discussed in Section~\ref{subsec:details}. In particular, we use the distilbert~\cite{distilbert}\footnote{https://huggingface.co/sentence-transformers/msmarco-distilbert-base-v3} as the retriever, and it consists of the 66M parameters. Also, for the reranker, we use the MiniLM model~\cite{minilm}\footnote{https://huggingface.co/cross-encoder/ms-marco-MiniLM-L-6-v2}, which consists of the 22M parameters. For supervised learning experiments, we train all models for 30 epochs, with a batch size of 512 for question answering and 32 for dialogue, and a learning rate of 2e-5. Also, we optimize all models using an AdamW optimizer~\cite{adamw}. We implement all models based on the following deep learning libraries: PyTorch~\cite{pytorch}, Transformers~\cite{transformers}, SentenceTransformers~\cite{sentence-bert}, and BEIR~\cite{beir}. For computing resources, we train and run all models with four GeForce RTX 2080 Ti GPUs and with Intel(R) Xeon(R) Gold 6240 CPU @ 2.60GHz having 72 processors. Also, training of our DiFaR framework takes less than one day. Note that we report all results with the single run, since our DiFaR framework significantly outperforms all baselines, but also it is costly to conduct multiple run experiments in the information retrieval experiment setting. 

\newpage

\section{Additional Experimental Results}
Here we provide additional experimental results.

\subsection{Running Time Efficiency}
Note that, while we provide running time comparisons between our DiFaR and DiFaR$^2$ in Figure~\ref{fig:efficiency}, it might be interesting to see more detailed running costs required for our dense fact retriever. As described in the Inference paragraph of Section~\ref{subsec:retrieval}, we index dense vectors with the Faiss library~\cite{faiss} that supports vector quantization and clustering for highly efficient search. Specifically, following the common vector index setting in previous document retrieval work~\cite{dpr, phrase}, we use the HNSW index type. Please refer to the documentation of the Faiss library\footnote{https://github.com/facebookresearch/faiss}\footnote{https://github.com/facebookresearch/faiss/wiki/Indexing-1M-vectors}, if you want to further explore different index types and their benchmark performances. 

We report running time efficiencies on the OpenDialKG dataset, which are measured on the server with Intel(R) Xeon(R) Gold 6240 CPU @ 2.60GHz having 72 processors (See Section~\ref{subsec:detail}). First of all, during inference, we can process about 174 queries per second where we return the top 1,000 facts for each query. Also, the average time for encoding and indexing one fact takes about 1 ms, which can be not only boosted further with more parallelization but also done in an online manner. Lastly, the performance drop of the approximation search with Faiss from the exact search is only 0.0098 on MRR.

\subsection{Additional Retrieval Examples}
\label{appendix:examples}
In this subsection, on top of the retrieval examples provided in Table~\ref{tab:example}, we provide the additional examples of our DiFaR framework in Table~\ref{tab:examples_additional}.

\onecolumn

\begin{center}

\begingroup

\fontsize{10pt}{12pt}\selectfont

\begin{longtable}{lp{1.5in}p{1.0in}p{1.5in}p{1.0in}}
\caption{\textbf{Retrieval examples of our DiFaR$^2$} on the Mintaka dataset for both single- and multi-hop questions.} 
\vspace{-0.1in}
\label{tab:examples_additional} \\

\hline 
\textbf{Index} & \textbf{Question} & \textbf{Question Entities} & \textbf{Retrieved Fact} & \textbf{Answer Entity} 
\\ \hline 
\endfirsthead

\multicolumn{5}{c}%
{{\bfseries \tablename\ \thetable{} -- Continued from the previous page}} \\
\hline 
\textbf{Index} & \textbf{Question} & \textbf{Question Entities} & \textbf{Retrieved Fact} & \textbf{Answer Entity} 
\\ \hline 
\endhead 

\multicolumn{5}{r}{\textit{\textbf{Continued on the next page}}} \\ \hline
\endfoot

\hline \hline
\endlastfoot

%%%%%%%%%%%%%%%%%%%%%%%%%%%%%% Examples %%%%%%%%%%%%%%%%%%%%%%%%%%%%%%%%%%%
1&
Which a series of unfortunate events books were not published in the 2000s?&
A Series of Unfortunate Events&
(A Series of Unfortunate Events, has part, The Bad Beginning)&
The Bad Beginnin\\

\noalign{\vskip 0.25ex}\cdashline{1-5}\noalign{\vskip 0.75ex}

2&
Who was the last leader of the soviet union?&
Soviet Union&
(Soviet Union, head of state, Mikhail Gorbachev)&
Mikhail Gorbachev\\

\noalign{\vskip 0.25ex}\cdashline{1-5}\noalign{\vskip 0.75ex}

3&
Who was the only u.s. vice president who is not male?&
U.S. vice president&
(Vice President of the United States, officeholder, Kamala Harris)&
Kamala Harris\\

\noalign{\vskip 0.25ex}\cdashline{1-5}\noalign{\vskip 0.75ex}

4&
Which author has won the most national book awards for fiction?&
National Book Awards for Fiction&
(National Book Award for Fiction, winner, Saul Bellow)&
Saul Bellow\\

\noalign{\vskip 0.25ex}\cdashline{1-5}\noalign{\vskip 0.75ex}

5&
Angkor wat can be found in which country?&
Angkor Wat&
(Angkor Wat, country, Cambodia)&
Cambodia\\

\noalign{\vskip 0.25ex}\cdashline{1-5}\noalign{\vskip 0.75ex}

6&
Albany is the capital of what state?&
Albany&
(Albany, capital of, New York)&
New York\\

\noalign{\vskip 0.25ex}\cdashline{1-5}\noalign{\vskip 0.75ex}

7&
Which u.s. president served the longest in office?&
U.S.&
(United States of America, head of government, Franklin Delano Roosevelt)&
Franklin Delano Roosevelt\\

\noalign{\vskip 0.25ex}\cdashline{1-5}\noalign{\vskip 0.75ex}

8&
Which state has the four largest cities in the united states and also does not share any borders with any other u.s. states?&
United States&
(United States of America, contains administrative territorial entity, Alaska)&
Alaska\\

\noalign{\vskip 0.25ex}\cdashline{1-5}\noalign{\vskip 0.75ex}

9&
What man was a famous american author and also a steamboat pilot on the mississippi river?&
Mississippi River, American&
(Life on the Mississippi, author, Mark Twain)&
Mark Twain\\

\noalign{\vskip 0.25ex}\cdashline{1-5}\noalign{\vskip 0.75ex}

10&
What country participated in ww ii and also used nuclear weapons in combat?&
WW II&
(Allies of the Second World War, has part, United States of America)&
United States of America\\

%%%%%%%%%%%%%%%%%%%%%%%%%%%%%%%%%%%%%%%%%%%%%%%%%%%%%%%%%%
\end{longtable}

\endgroup

\end{center}

\twocolumn

\end{document}